\documentclass{emulateapj}

\countdef\decade=200
\advance\decade by \year
\countdef\hours=201
\advance\hours by \time
\divide\hours by 60
\countdef\mins=202
\advance\mins by \hours
\multiply\mins by 60
\multiply\hours by 100
\countdef\miltime=203
\advance\miltime by \hours
\advance\miltime by \time
\advance\miltime by -\mins

\newcommand{\ltsim}{\protect\raisebox{-0.5ex}{$\:\stackrel{\textstyle <}
        {\sim}\:$}}
\newcommand{\gtsim}{\protect\raisebox{-0.5ex}{$\:\stackrel{\textstyle >}
        {\sim}\:$}}

\newcommand{\lsun}{L_{\odot}}

\newcommand{\nc}{n_{\rm crit}}
\newcommand{\nbar}{\overline{n}}
\newcommand{\nmed}{n_{\rm med}}
\newcommand{\tff}{t_{\rm ff}}
\newcommand{\sfrff}{\mbox{SFR}_{\rm ff}}
\newcommand{\calm}{\mathcal{M}}
\newcommand{\avir}{\alpha_{\rm vir}}
\newcommand{\mcl}{M_{\rm cl}}
\newcommand{\jline}{1\rightarrow 0}

\shorttitle{Molecular Gas and Kennicutt-Schmidt Laws}

\begin{document}

\title{The Relationship Between Molecular Gas Tracers and
Kennicutt-Schmidt Laws}

\slugcomment{Accepted for publication in the Astrophysical Journal, July 16, 2007}

\author{Mark R. Krumholz\altaffilmark{1} and 
Todd A.~Thompson\altaffilmark{2}}
\affil{Department of Astrophysical Sciences, Princeton University,
Princeton, NJ 08544}
\email{krumholz@astro.princeton.edu, thomp@astro.princeton.edu}

\altaffiltext{1}{Hubble Fellow}
\altaffiltext{2}{Lyman Spitzer Jr.~Fellow}

\begin{abstract}
We provide a model for how Kennicutt-Schmidt (KS) laws, which describe the
correlation between star formation rate and gas surface or volume
density, depend on the molecular line chosen to trace the
gas. We show that, for lines that can be excited at low temperatures,
the KS law depends on how the line critical density compares to
the median density in a galaxy's star-forming molecular clouds.
High critical density lines trace regions with similar
physical properties across galaxy types,
and this produces a linear correlation between line luminosity and
star formation rate. Low critical density lines probe regions
whose properties vary across galaxies, leading to a star formation
rate that varies superlinearly with line luminosity. We show that a
simple model in which molecular clouds are treated as isothermal and
homogenous can quantitatively reproduce the observed correlations
between
galactic luminosities in far infrared and in the CO($\jline$) and
HCN($\jline$) lines, and naturally explains why these correlations
have different slopes. We predict that IR-line luminosity correlations
should change slope for galaxies in which the median density is close
to the line critical density. This prediction may be tested by
observations of lines such as HCO$^+$($\jline$) with intermediate
critical densities, or by HCN($\jline$) observations of intensely
star-forming high redshift galaxies with very high densities. Recent
observations by Gao et al.\ hint at just such a change in slope.
We argue that deviations from linearity in the HCN($\jline$)$-$IR correlation at high luminosity
are consistent with the assumption of a constant star formation 
efficiency.
\end{abstract}

\keywords{ISM: clouds --- ISM: molecules --- stars: formation ---
galaxies: ISM --- radio lines: ISM}

\section{Introduction}
\label{intro}

\citet{schmidt59,schmidt63} first proposed that the rate at which a
gas forms stars might follow a simple power law correlation of the
form $\dot{\rho}_* \propto \rho_g^N$, where $\dot{\rho}_*$ is the
star formation rate per unit volume, $\rho_g$ is the gas
density, and $N$ is generally taken to be in the range $1-2$. 
In the decades since, observations have revealed two strong
correlations that appear to be evidence for this hypothesis. First,
galaxy surveys reveal that
the infrared luminosity of a galaxy, which traces the star formation
rate, varies with its luminosity in the CO($\jline$) line, which
traces the total mass of molecular gas, as $L_{\rm FIR}\propto L_{\rm
CO}^{1.4-1.6}$
\citep{gao04b,gao04a,greve05,riechers06a}. \citet{kennicutt98b,
kennicutt98a} identified the closely-related correlation between gas
surface density $\Sigma_g$ and star formation rate surface
density $\dot{\Sigma}_*$, $\dot{\Sigma}_*\propto \Sigma_g^{1.4\pm
0.15}$, a relation that has come to be known as the Kennicutt
Law. Since over the bulk of the dynamic range of Kennicutt's data
galaxies are predominantly molecular, this is effectively a
correlation between molecular gas, as traced by CO($\jline$) line
emission, and star formation. Spatially resolved observations of 
galaxies confirm that, at least for molecule-rich galaxies where 
resolved CO($\jline$) observations are possible, star formation is 
more closely coupled with gas traced by CO($\jline$) than with 
atomic gas \citep{wong02,heyer04b,komugi05,kennicutt07a}

Second, \citet{gao04b,gao04a} find that there is a strong 
correlation between the IR luminosity of galaxies and emission in
the HCN($\jline$) line, which measures the mass at densities
significantly greater than that probed by CO($\jline$). However, they
find that their correlation, which covers nearly three decades in
total galactic star formation rate, is linear: $L_{\rm FIR} \propto
L_{\rm HCN}$. \citet{wu05} show that this correlation extends down to
individual star-forming clumps of gas in the Milky Way, provided that
their infrared luminosities are $\gtsim 10^{4.5}$
$\lsun$. Interestingly, however, \citet{gao07a} find a deviation from
linearity in the IR-HCN correlation for a sample of
intensely star-forming high redshift galaxies. These sources show
small but significant excesses of infrared emission for their
observed HCN emission.

The difference in power law indices between the $L_{\rm FIR}-L_{\rm CO}$
and $L_{\rm FIR}- L_{\rm HCN}$ correlations is
statistically significant, and, on its face, puzzling. An index near
$N=1.5$ 
seems natural if one supposes that a roughly constant fraction of the gas
present in molecular clouds will be converted into stars each
free-fall time. In this case one expects $\dot{\rho}_* \propto
\rho_g^{1.5}$ \citep{madore77, elmegreen94b}. If gas scale heights do
not vary strongly from galaxy to galaxy, this implies
$\dot{\Sigma}_*\propto \Sigma_g^{1.5}$ as well, which is consistent
with the observed Kennicutt law. More generally, since the
dynamical timescale in a marginally Toomre-stable 
\citep[$Q\approx1$; see][]{martin01}
galactic disk is of order $\Omega^{-1}\propto(G\rho_g)^{-1/2}$,
where $\Omega$ is the angular frequency of the disk,
an index close to $N=1.5$ is expected if star formation is regulated
by any phenomenon that converts a fixed fraction of the gas into stars
on this time scale \citep{elmegreen02}. 

On the other hand, \citet{wu05} suggest a simple interpretation of the
linear IR-HCN correlation. They argue that the individual
HCN-emitting molecular clumps that they identify in the Milky Way
represent a fundamental unit of star formation. The linear correlation
between star formation rate and HCN luminosity across galaxies arises
because a measurement of the HCN luminosity for a galaxy simply counts
the number of such structures present within it, each of
which forms stars at some roughly fixed rate regardless of its
galactic environment. However, in this interpretation it is unclear
why the structures traced by HCN$(\jline$) emission should
form stars at the same rate in any galaxy. After all, one could
equally well argue that molecular clouds traced by CO($\jline$)
are fundamental units of star formation, but the non-linear
IR-CO correlation clearly shows that
these objects do not form stars at a fixed rate per unit
mass. Moreover, the evidence presented by \citet{gao07a} that
the linear IR-HCN correlation varies in extremely luminous high
redshift galaxies suggests that the relationship between HCN emission
and star-formation may be somewhat more complex.

In this paper we attempt to explain the origin of the difference in
slope between the CO and HCN correlations with star formation rate,
and more generally to give a theoretical framework for understanding
how correlations between star formation rate and line luminosity,
which we generically refer to as Kennicutt-Schmidt (KS) laws, depend
on the tracer used to define them. Our central argument is
conceptually quite simple, and in some sense represents a combination
of the intuitive arguments for CO and HCN given above.

Consider an observation of a galaxy in a
molecular tracer with critical density $\nc$, which essentially
measures the mass of gas at densities of $\nc$ or more, i.e.\ the gas
that is dense enough for that particular transition to be excited.
In galaxies where the median density of the molecular gas is
significantly larger than
$\nc$, this means that the observation will detect the majority of the
gas, and the bulk of the emission will come from gas whose density is
near the median density. Since
the gas density will vary from galaxy
to galaxy, the star formation rate per unit gas mass will vary as
roughly $\rho_g^{1.5}$, with one factor of $\rho_g$ coming from the amount
of gas available for star formation, and an additional factor of
$\rho_g^{0.5}$ coming from the dependence of the free-fall or
dynamical time on the density.

On the other hand, in galaxies where the median gas density
is small compared to the critical density for the chosen transition,
observations will pick out only high density peaks. Since the density
in these peaks is set by $\nc$, and not by the conditions in the
galaxy, these peaks are at essentially the same density in any galaxy
where they are observed, and the corresponding free-fall times in
these regions are constant as well. As a result, the star formation
rate per unit mass of gas traced by that line is approximately the
same in every galaxy, because the corresponding free-fall time is the
same in every galaxy. 

In the rest of this paper, we give a quantitative version of this
intuitive argument, and then discuss its consequences. In
\S~\ref{formalism} we develop a simple formalism to compute the star
formation rate and the molecular line luminosity of galaxies, and in
\S~\ref{predictions} we use this formalism to predict the correlation
between star formation rate and luminosity. We show that our
predictions provide a very good fit for a variety of observations, and
make predictions for future observations. We discuss the implications
of our work and its limitations in \S~\ref{discussion}, and summarize
our conclusions is \S~\ref{conclusions}.

\section{Star Formation Rates and Line Luminosities}
\label{formalism}

\subsection{Cloud Properties}

Consider a galaxy in which the star-forming molecular clouds have a
volume-averaged mean molecular hydrogen number density $\nbar =
\overline{\rho}_g/\mu_{H_2}$,\ where $\overline{\rho}_g$ is the volume-averaged mass density of the molecular clouds in the galaxy and $\mu_{H_2}=3.9\times 10^{-24}$ g is
the mean mass per hydrogen molecule for a gas of standard cosmic
composition. Observations indicate that
$\nbar$ varies by two to three decades over the galaxies for which
the Kennicutt and Gao \& Solomon correlations are measured, from
$\nbar\approx 50$ cm$^{-3}$ in normal spirals like the Milky Way
\citep{mckee99a} up to $\nbar \approx 10^4$ cm$^{-3}$ in the strongest
starburst systems in the local universe \citep[e.g.][]{downes98}.
There is strong evidence that
densities in molecular clouds follow a lognormal probability
distribution function (PDF; see reviews by \citealt{maclow04} and
\citealt{elmegreen04})
\begin{equation}
\frac{dp}{d\ln x} = \frac{1}{\sqrt{2\pi\sigma^2}}
\exp\left[-\frac{(\ln x - \overline{\ln x})^2}{2\sigma^2}\right],
\end{equation}
where $x=n/\nbar$ is the molecular hydrogen number density $n$
relative to the average density, $\sigma$ is the
width of the lognormal, and $\overline{\ln x} = -\sigma^2/2$. For this
distribution the median density is $n_{\rm med} =
\nbar\exp(\sigma^2/2)$. Numerical experiments show that for supersonic
isothermal turbulence $\sigma^2\approx \ln\left(1+3 \calm^2/4\right)$,
where $\calm$ is the 1D Mach number of the turbulence
\citep{nordlund99, ostriker99, padoan02}. Mach numbers in
star-forming molecular clouds range from $\calm \sim 30$
\citep{mckee99a} in normal spirals to $\calm\sim 100$ in strong
starbursts \citep{downes98}, implying that median densities in
molecular clouds range from $\sim 10^3$ cm$^{-3}$ in normal spirals to
$\sim 10^6$ cm$^{-3}$ in starbursts. Star forming clouds within a
galaxy are approximately isothermal, except very near strong sources
of stellar radiation, so we assume a fixed temperature $T$ for the
clouds. Observationally, $T$ ranges from roughly 10 K in normal
spirals \citep{mckee99a} up to as much about 50 K in strong starbursts
\citep{downes98, gao04a}.

\subsection{Star Formation Rates}

First let us ask how quickly stars form in such a
medium. \citet{krumholz05c} give a model for star formation regulated
by supersonic turbulence in which a population of molecular clouds of
total mass $\mcl$ form stars at a rate $\dot{M}_* = \sfrff \mcl /
\tff(\nbar)$, where $\tff(\nbar)$ is the free-fall time evaluated at
the mean density and $\sfrff$ is a number of order $10^{-2}$ that
depends weakly on $\calm$. We therefore estimate the star formation
rate per unit volume as a function of the mean density given by
\begin{equation}
\label{SFR}
\dot{\rho}_* \approx \sfrff \sqrt{\frac{32 G \mu_{H_2}^3 \nbar^3}{3\pi}}.
\end{equation}
We adopt the Krumholz \& McKee result $\sfrff\approx 0.014
(\calm/100)^{-0.32}$ for clouds with a fiducial virial ratio of
$\avir=1.3$.

Alternately, \citet{krumholz07e} point out that observed
correlations between the star formation rate and the luminosity in
different density tracers imply that over a $3-4$ decade range in
density $n$,
\begin{equation}
\label{SFR1}
\dot{M}_* \approx 10^{-2} \frac{\mcl(>n)}{\tff(n)},
\end{equation}
where $\mcl(>n)$ is the mass of gas with a density of $n$ or
higher, and $\mcl=\mcl(>0)$.
For a given choice of $n$ this provides an alternative
estimate of the star formation
rate which is purely empirical, and independent of any particular
theoretical model. However, the difference between the star formation
rates predicted by (\ref{SFR}) and (\ref{SFR1}) is small.
For gas with a lognormal PDF,
\begin{equation}
\mcl(>n) = \frac{\mcl}{2} \left(1 + 
\mbox{erf}\left[\frac{-2 \ln x+\sigma^2}{2^{3/2}\sigma}
\right]\right),
\end{equation}
and using this to evaluate equation (\ref{SFR1}) indicates that, for
Mach numbers in the observed range, the two prescriptions (\ref{SFR})
and (\ref{SFR1}) give about the same star formation
rate over a very broad range in $x$. For example, at $\calm=30$ the two
estimates agree to within a factor of $3$ for densities in the range 
$0.2 < x < 4\times 10^{4}$. Given the scatter inherent in
observational estimates of the star formation rate, a factor of 3
difference is not particularly significant, so it matters little which
prescription we adopt. In practice, we will use equation (\ref{SFR}).

\subsection{Line Luminosities}

Now we must compute the luminosity of molecular line emission from the
galaxy. Even for a cloud that is not in local thermodynamic
equilibrium (LTE), for optically thin emission this calculation is
straightforward. However, the molecular lines used most often in
galaxy surveys are generally optically thick. To handle the effect of
finite optical depth on molecule level populations and line
luminosities, we adopt an escape probability approximation and treat
clouds as homogeneous spheres. This is not fully consistent with our
assumption that clouds have lognormal density PDFs,
since the escape probability formalism assumes a uniform level
population throughout the cloud, and the essence of our argument in
this paper turns on how the level population varies with
density. However, this approach gives us an approximate way of
incorporating the optical thickness of star-forming clouds into our
model, the only alternative to which for turbulent media is full
numerical simulation \citep[e.g.][]{juvela01}. We therefore proceed by
treating clouds as homogeneous in order to determine their escape
probabilities, and we then relax the assumption of homogeneity, while
keeping the escape probabilities fixed, in order to determine level
populations and cloud luminosities as a function of density.

Consider a cloud of radius $R$ in statistical equilibrium but not
necessarily in LTE. In the
escape probability approximation, the fraction $f_i$ of molecules of
species $S$ in state $i$ is given implicitly by the linear system
\begin{eqnarray}
\label{levpop}
\sum_j \left(n q_{ji} + \beta_{ji} A_{ji}\right) f_j & = &
\left[\sum_j\left(n q_{ij} + \beta_{ij} A_{ij}\right)\right] f_i \\
\label{levpop1}
\sum_i f_i & = & 1,
\end{eqnarray}
where $q_{ij}$ is the collision rate for transitions from state $i$ to
state $j$, $A_{ij}$ is the Einstein spontaneous emission
coefficient for this transition, $\beta_{ij}$ is the cloud-averaged
escape probability for photons emitted in this transition, the sums
are over all quantum states, and we understand that $A_{ij} = 0$ for
$i\leq j$ and $q_{ij} = 0$ for $i=j$.

Equations (\ref{levpop}) and (\ref{levpop1}) allow us to compute the level
populations $f_i$ for given values of $\beta_{ij}$. To completely
specify the system, we must add an additional consistency condition
relating the values of $\beta_{ij}$ to the level populations. For a
homogeneous spherical cloud, the escape probability for a given line is
related to the optical depth from the center to the edge of the cloud
$\tau_{ij}$ by (B.~Draine, 2007, private communication)
\begin{equation}
\label{taucondition1}
\beta_{ij} \approx \frac{1}{1+0.5\tau_{ij}},
\end{equation}
where $\tau_{ij}$ is computed at the central frequency of the line. In
turn, the optical depth is related to the level populations by 
\begin{equation}
\label{taucondition2}
\tau_{ij} = \frac{g_j}{g_j} \frac{A_{ij}
\lambda_{ij}^3}{4 (2\pi)^{3/2} \calm c_s} \nbar X(S) f_j R
\left(1-\frac{f_i g_j}{f_j g_i}\right),
\end{equation}
where $\lambda_{ij}$ is the wavelength of transition $i\rightarrow j$,
$g_i$ and $g_j$ are the statistical weights of
states $i$ and $j$, $c_s$ is the isothermal sound speed of the gas, and
$X(S)$ is the abundance of molecules of species $S$. Note that this
equation implicitly assumes that the cloud has a uniform Maxwellian
velocity distribution with 1D dispersion $\calm c_s$, consistent with
our treatment of the clouds as homogeneous spheres.
One additional complication is that we do not directly know cloud
radii for most external galaxies, where observations cannot resolve
individual molecular clouds. However, we often can diagnose the optical
depths of transitions by comparing line ratios of molecular
isotopomers of different abundances. We therefore take $\tau_{10}$, the
optical depth of the transition between the first excited state and
the ground state, as known. For a given level population this fixes
the value of $R$.

We solve equations (\ref{levpop})--(\ref{taucondition2}) using
Newton-Raphson iteration. In this procedure, we guess an initial
set of escape probabilities $\beta_{ij}$, and solve the linear system
(\ref{levpop}) and (\ref{levpop1}) to find the corresponding initial
level populations $f_i$. We then compute the optical
depths $\tau_{ij}$ from equation (\ref{taucondition2}). The
guessed escape probabilities $\beta_{ij}$ and the corresponding
optical depths $\tau_{ij}$ generally will not satisfy
the consistency condition  (\ref{taucondition1}), so we then iterate
over $\beta_{ij}$ values using a Newton-Raphson approach, seeking
$\beta_{ij}$ for
which the level populations give optical depths $\tau_{ij}$ such that
all elements of the matrix $\beta_{ij} - 1/(1+0.5\tau_{ij})$ are equal
to zero within some specified tolerance. We use the LTE level
populations and escape probabilities for our initial guess, so that
the iteration converges rapidly when the system is close to LTE.

\begin{deluxetable*}{ccccc}
\tablecaption{Model Parameters
\label{fiducial}}
\tablewidth{0pt}
\tablehead{
\colhead{Parameter} &
\colhead{Normal galaxy} &
\colhead{Intermediate} &
\colhead{Starburst} &
\colhead{Reference}
}
\startdata
T & 10 & 20 & 50 & 1--4 \\
$\calm$ & 30 & 50 & 80 & 1--4 \\
X(CO) & $2\times 10^{-4}$ & $4\times 10^{-4}$ & $8\times 10^{-4}$ & 5\\
X(HCO$^+$) & $2\times 10^{-9}$ & $4\times 10^{-9}$ & $8\times 10^{-9}$ & 6, 7 \\
X(HCN) & $1\times 10^{-8}$ & $2\times 10^{-8}$ & $4\times 10^{-8}$ & 6--8 \\
$\tau_{{\rm CO}(\jline)}$ & 10 & 20 & 40 & 9 \\
$\tau_{{\rm HCO}^+(\jline)}$ & 0.5 & 1.0 & 2.0 & 6, 7 \\
$\tau_{{\rm HCN}(\jline)}$ & 0.5 & 1.0 & 2.0 & 6, 7 \\
OPR & 0.25 & 0.25 & 0.25 & 10\\
\enddata
\tablecomments{OPR = H$_2$ ortho- to para-ratio.
References: 1 -- \citet{solomon87}, 2 -- \citet{gao04a}, 3 --
\citet{downes98}, 4 -- \citet{wu05}, 5 -- \citet{black00}, 
6 -- \citet{nguyen92}, 7 -- \citet{wild92}, 8 -- \citet{lahuis00},
9 -- \citet{combes91}, 10 -- \citet{neufeld06}
}
\end{deluxetable*}

Once we have determined the escape probabilities $\beta_{ij}$, we
compute the luminosity by holding the $\beta_{ij}$ values fixed but
allowing the level populations to vary with density, then integrating
over the PDF. Thus, the total luminosity per unit volume in a
particular line is
\begin{equation}
\label{linelum}
L_{ij} = X(S) \beta_{ij} A_{ij} h \nu_{ij}
\int_{-\infty}^{\infty} f_i n \frac{dp}{d\ln x} d\ln x,
\end{equation}
where $\nu_{ij}$ is the line frequency, $f_i$ is an implicit function
of $n$ given by the solution to equations (\ref{levpop}) and
(\ref{levpop1}), and we assume that the abundance $X(S)$ is
independent of $n$. The line
luminosity per unit mass is $L_{ij}/(\mu_{H_2} \nbar)$.

An IDL code that implements this calculation is available for public download from http://www.astro.princeton.edu/$\sim$krumholz/ astronomy.html.

\section{Correlations and Kennicutt-Schmidt Laws}
\label{predictions}

\subsection{Lines and Parameters}

Using the formalism of \S~\ref{formalism}, we can now predict the
correlation between the star formation rate and the luminosity of a
galaxy in molecular lines. We make these predictions for three
representative molecular lines: CO($\jline$),
HCO$^+$($\jline$), and HCN($\jline$). For the first and
last of these transitions, there are extensive observational
surveys. We select HCO$^+$($\jline$) in addition to these
two because there is some observational data for it, and because
its critical density of $\nc=\beta_{\rm HCO+} 4.6\times
10^4$ cm$^{-3}$ makes it intermediate between CO($\jline$), with
$\nc=\beta_{\rm CO} 560$ cm$^{-3}$, and HCN($\jline$), with
$\nc=\beta_{\rm HCN} 2.8\times 10^5$ cm$^{-3}$.\footnote{Note that our
critical density for HCN($\jline$)
is somewhat larger than the value quoted by \citet{gao04b,gao04a},
probably because their calculation is based on somewhat different
assumptions about how to extrapolate from calculated rate coefficients
for HCN collisions with He to collisions with H$_2$. See
\citet{schoier05} for details.} Here $\beta_{S}$ is
the escape probability for the $\jline$ transition of species
S. These critical densities are for
$T=20$ K. All molecular data are taken from the Leiden Atomic and
Molecular
Database\footnote{http://www.strw.leidenuniv.nl/$\sim$moldata/}
\citep{schoier05}.

We make our calculations for three sets
of fiducial parameters which we summarize in Table
\ref{fiducial}. The three sets correspond roughly to typical
conditions in normal disk galaxies like the Milky Way, to starburst
galaxies like Arp 220, and to a case intermediate between the
two. We have selected parameters for each case to roughly model the
systematic variation of ISM parameters as one moves from normal disk
galaxies to starbursts. Thus, we vary the ISM temperature from $10-50$
K and the molecular cloud Mach number from $30-80$ as we move from
Milky Way-like molecular clouds to temperatures and Mach numbers
typical of starbursts
\citep[e.g.][]{downes98}. Similarly, starbursts, which preferentially
occur at galactic centers, have systematically larger metallicities
than galaxies like the Milky Way
\citep[e.g.][]{zaritsky94,yao03,netzer05}. To explore this effect, we
use abundances and $\jline$ optical depths are twice and four times as
large for our intermediate and starburst models, respectively, as for
our normal galaxy model.

\subsection{Kennicutt-Schmidt Laws}
\label{schmidtlaws}

\begin{figure}
\plotone{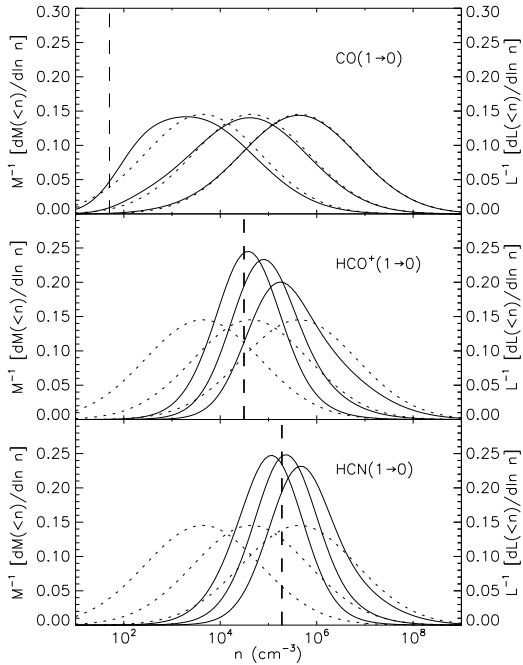}
\caption{\label{dldlogn}
Fractional contribution to the total luminosity $L^{-1}
[dL(<n)/d\ln n]$ (\textit{solid lines}) and mass $M^{-1}
[dM(<n)/d\ln n]$ (\textit{dotted lines}) 
versus density $n$ for the lines CO($\jline$) (\textit{top panel}), HCO$^+$($\jline$) (\textit{middle panel}), and HCN($\jline$) (\textit{bottom panel}). The three curves show the cases
$\nbar=10^2$ cm$^{-3}$, $10^3$ cm$^{-3}$, and $10^4$ cm$^{-3}$,
from leftmost to rightmost. We also show the critical density of each
molecule, corrected for radiative trapping (\textit{dashed
vertical lines}). These calculations use the parameters for the intermediate case listed in Table \ref{fiducial}.\\
}
\end{figure}

We first plot, in Figure \ref{dldlogn}, the quantities $L^{-1}
[dL(<n)/d\ln n]$ ({\it solid lines}) and $M^{-1}[dM(<n)/d\ln n]$ ({\it dotted lines})
as a function of density $n$ for galaxies with mean densities 
$\nbar=10^2$, $10^3$, and $10^4$ cm$^{-3}$, for the tracers
CO($\jline$), HCO$^+$($\jline$), and HCN($\jline$), and for the Mach number and temperature corresponding to our intermediate case in Table \ref{fiducial}.  
Here $L(<n)$ and $M(<n)$ are the luminosity and mass per unit volume 
contributed by gas of density $n$ or less, i.e.\ $L(<n)=X(S) \beta_{ij} A_{ij} h \nu_{ij}
\int_{-\infty}^{\ln n} f_i n (dp/d\ln n) d\ln n$, $M(<n) =
\int_{-\infty}^{\ln n} \mu_{H_2} n (dp/d\ln n) d\ln n$,
$L=L(<\infty)$,  and $M=M(<\infty)$. Physically, $L^{-1}[dL(<n)/d\ln n]$ and $M^{-1}[dM(<n)/d\ln n]$ represent the
fractional contribution to the total line luminosity and the total
mass that comes from each unit interval in the logarithm of
density. The plot shows what density range provides the dominant
contribution to the line luminosity in different lines and for
galaxies of differing mean densities, and how the gas contributing
light compares to the gas contributing mass. Because the mass distribution
is entirely specified by $\nbar$ and $\calm$, the dotted lines are the 
same in each of the three panels.  
Additionally, because of our choice $\calm=50$ 
(Table \ref{fiducial}), the median density (the density corresponding
to the peak in $M^{-1}[dM(<n)/d\ln n]$) is $n_{\rm med}\approx43\,\nbar$.
In each panel, the critical density for each molecule 
is identified by a vertical dashed line.
 
The top panel clearly shows that for the CO line, the light 
and the mass track one another very closely, even
at the lowest densities. Thus, because $\nmed>\nc$, 
the solid lines move in lock-step with the dashed lines as $\nbar$ increases.
In contrast, for HCN most of the luminosity
comes from densities near the critical density regardless of the mass
distribution. For the lowest $\nbar$ this means that the line luminosity
is entirely dominated by the high density tail of the mass distribution. 
As the median density $\nmed$ varies by a
factor of 100 (from $4.3\times10^3-4.3\times10^5$ cm$^{-3}$), 
the peak of $L^{-1} [dL(<n)/d\ln n]$ moves
by just a factor of a few in $n$. The HCO$^+$ line is
intermediate between CO and HCN. For $\nbar=10^2$ cm$^{-3}$ and
$10^3$ cm$^{-3}$, $\nmed\ltsim\nc$, and as with HCN most of the
emission comes from near the critical density. For $\nbar=10^4$
cm$^{-3}$, $\nmed>\nc$, and the light starts to follow the
mass, in a pattern similar to that for CO. Although Figure \ref{dldlogn} shows only the intermediate case, the normal galaxy and starburst cases give qualitatively identical results. This
confirms the intuitive argument given in \S~\ref{intro}: \textit{high
critical density transitions trace regions of similar density in every
galaxy, while low critical density transitions trace regions whose
density is close to the median density}.

Now consider how the luminosity in a given line correlates with the
star formation rate in galaxies of varying mean densities. For a given $\nbar$, we can compute
the volume density of star formation from equation (\ref{SFR}) and the
line luminosity density from (\ref{linelum}). To
facilitate comparison with observations, rather than considering the
total line luminosity, we use the quantity $L'$
\citep{solomon97}, which is related to the luminosity $L$ by
\begin{equation}
L' = \frac{c^2}{8\pi k_B \nu^2} L,
\end{equation}
converted to the units K km s$^{-1}$ pc$^{2}$.

Similarly, we can estimate the far infrared luminosity from the 
star formation rate. There is a tight correlation between far-IR 
emission and star formation, particularly for dense, dusty galaxies 
like those that make up most of the dynamic range of the 
\citet{kennicutt98b} sample. To the extent that most or all of the 
light from young stars is re-processed by dust before escaping the 
galaxy, the bolometric luminosity integrated over the wavelength 
range $8-1000$ $\mu$m, which we define as $L_{\rm FIR}$, simply
provides a calorimetric measurement of the total energy output by
young stars, and is therefore an excellent tracer of recent star formation 
\citep{sanders96, rowanrobinson97, kennicutt98b, kennicutt98a,
hirashita03,bell03,iglesiasparamo06}. We therefore estimate the FIR
luminosity from the star formation rate via
\begin{equation}
\label{sfrtolir}
L_{\rm FIR} = \epsilon \dot{M}_* c^2,
\end{equation}
where $\epsilon$ is an IMF-dependent constant. For consistency with
\citet{kennicutt98b,kennicutt98a}, we take $\epsilon=3.8\times 10^{-4}$. 
To be precise and to facilitate comparison with observations, we adopt the 
\citet{sanders96} definition of $L_{\rm FIR}$ as a weighted sum of the 
luminosity in the 60 and 100 $\mu$m IRAS bands.  This definition
of the infrared luminosity generically underestimates the total 
infrared luminosity $[8-1000]\mu$m by a factor of $1.5-2$
\citep{calzetti00, dale01, bell03}. However, we use the $\epsilon$
value appropriate for $L_{\rm FIR}$ rather than for the total IR
luminosity because some of the observations to which we wish to
compare our model (see \S~\ref{obscomparison}) provide only $L_{\rm
FIR}$.
Note that this choice for the connection between the star formation rate 
and the infrared luminosity is not fully consistent with our choice of the 
gas temperature for the three sets of parameters ---
normal, intermediate, and starburst --- listed in Table \ref{fiducial}, 
an issue we discuss in more detail in \S~\ref{limitations}.

\begin{figure}
\plotone{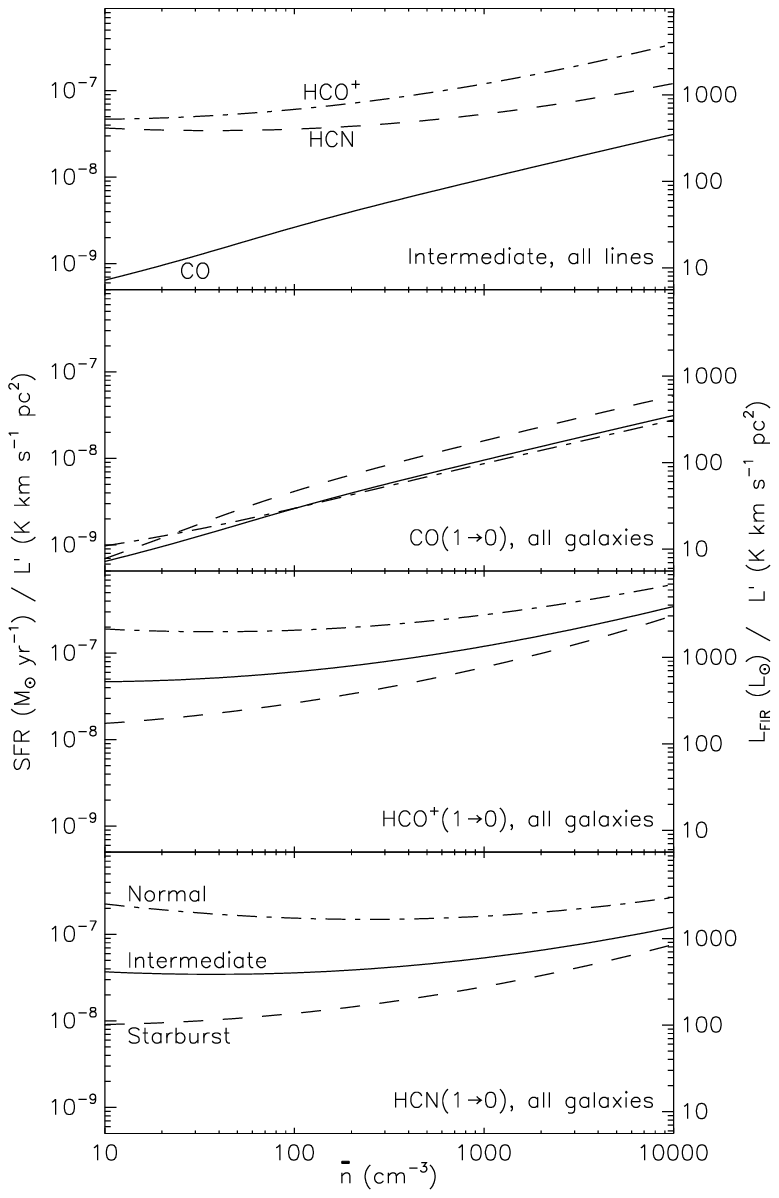}
\caption{
\label{sfrl}
Ratio of star formation rate or infrared luminosity to line luminosity,
as a function of
mean density $\nbar$. In the top panel we show the lines CO($\jline$) (\textit{solid
line}), HCO$^+$($\jline$) (\textit{dot-dashed line}), and
HCN($\jline$) (\textit{dashed line}) for the intermediate case in Table \ref{fiducial}. In the next three panels we show the CO($\jline$), HCO$^+$($\jline$), and HCN($\jline$) lines for the normal galaxy case (\textit{dot-dashed line}), intermediate case (\textit{solid line}), and starburst case (\textit{dashed line}).
}
\end{figure}

We plot the ratio of star formation rate to line luminosity, and infrared 
luminosity to line luminosity, as a function of $\nbar$ in Figure \ref{sfrl}. 
First consider the top panel, which shows all three lines computed for the 
intermediate case. This again confirms our intuitive argument. Since the 
luminosity per unit volume in the CO line is roughly proportional to the 
mass density, and the star formation rate / IR luminosity is proportional 
to mass density to the 1.5 power, the
ratios $\dot{M}_*/L'$ and $L_{\rm FIR}/L'$ vary roughly
as $\nbar^{0.5}$. A powerlaw fit to
the data over the range shown in Figure \ref{sfrl} gives an index of
0.57. In contrast, the ratio of star formation density to HCN
luminosity density is nearly constant for galaxies with $\nbar<10^3$
cm$^{-3}$, and varies quite weakly with $\nbar$ up to densities of
$10^4$ cm$^{-3}$, values found in the densest starbursts. A powerlaw
fit from $10$ cm$^{-3}$ to $10^4$ cm$^{-3}$ gives an index of 0.17;
from $10$ cm$^{-3}$ and $10^3$ cm$^{-3}$, the best fit powerlaw index
is $0.08$. As
in Figure \ref{dldlogn}, the slope of the $\dot{M}_*/L'$ curve for HCO$^+$ 
represents an intermediate case, with a roughly constant ratio of $\dot{M}_*/L'$ 
and $L_{\rm FIR}/L'$
at low $\nbar$, rising to a
slope comparable to that for CO at high values of $\nbar$.

Now consider the bottom three panels in Figure \ref{sfrl}. 
Each panel shows the ratio of
star formation rate and infrared luminosity to line luminosity for a
single line, computed for each of the three galaxy models. The most
important point to take from these plots is that the choice of galaxy
model has little effect in most cases. The largest differences are for
HCN, where at $\nbar=10$ cm$^{-3}$ the IR to line ratio predicted for the
intermediate case differs from the normal galaxy case by a factor of
$6.1$, and from the starburst case by a factor of $4.1$. This
variation comes primarily from changes in the Mach number and the
optical depth between models. The higher Mach number of the starburst
model significantly increases the amount of mass in the high
overdensity tail of the probability distribution, while the higher
optical depth lowers the effective critical density. Both of these
effect increase the amount of mass dense enough to emit in
HCN($\jline$) and reduce $\dot{M}_*/L'$. At higher mean densities
these effects become less important and the models converge, so that
by $\nbar=10^4$ cm$^{-3}$ the range in $\dot{M}_*/L'$ from the normal
to the starburst case is only a factor of 3.5.

Most importantly, our central conclusion that $\dot{M}_*/L'_{\rm HCN}$
is roughly constant across galaxies,  while $\dot{M}_*/L'_{\rm CO}$
rises as roughly $[L'_{\rm CO}]^{0.5}$, still holds when we consider
how conditions vary across galaxies. Galaxies with low mean densities
$\nbar$ are generally closest to the normal galaxy case, while those
with high mean densities should be closest to the starburst case, and
this systematic variation in galaxy properties with $\nbar$ still
leaves $\dot{M}_*/L'$ relatively flat for HCN, and varying with a
slope close to 0.5 for CO. From the normal galaxy case at $\nbar=10$
cm$^{-3}$ to the starburst case at $\nbar=10^4$ cm$^{-3}$, the value
of $\dot{M}_*/L'$ varies by more than a factor of 50 for the
CO($\jline$), but by less than a factor of 3 for the HCN($\jline$).

\subsection{Comparison with Observations}
\label{obscomparison}

\begin{figure*}
\centerline{
\includegraphics[width=5.25cm]{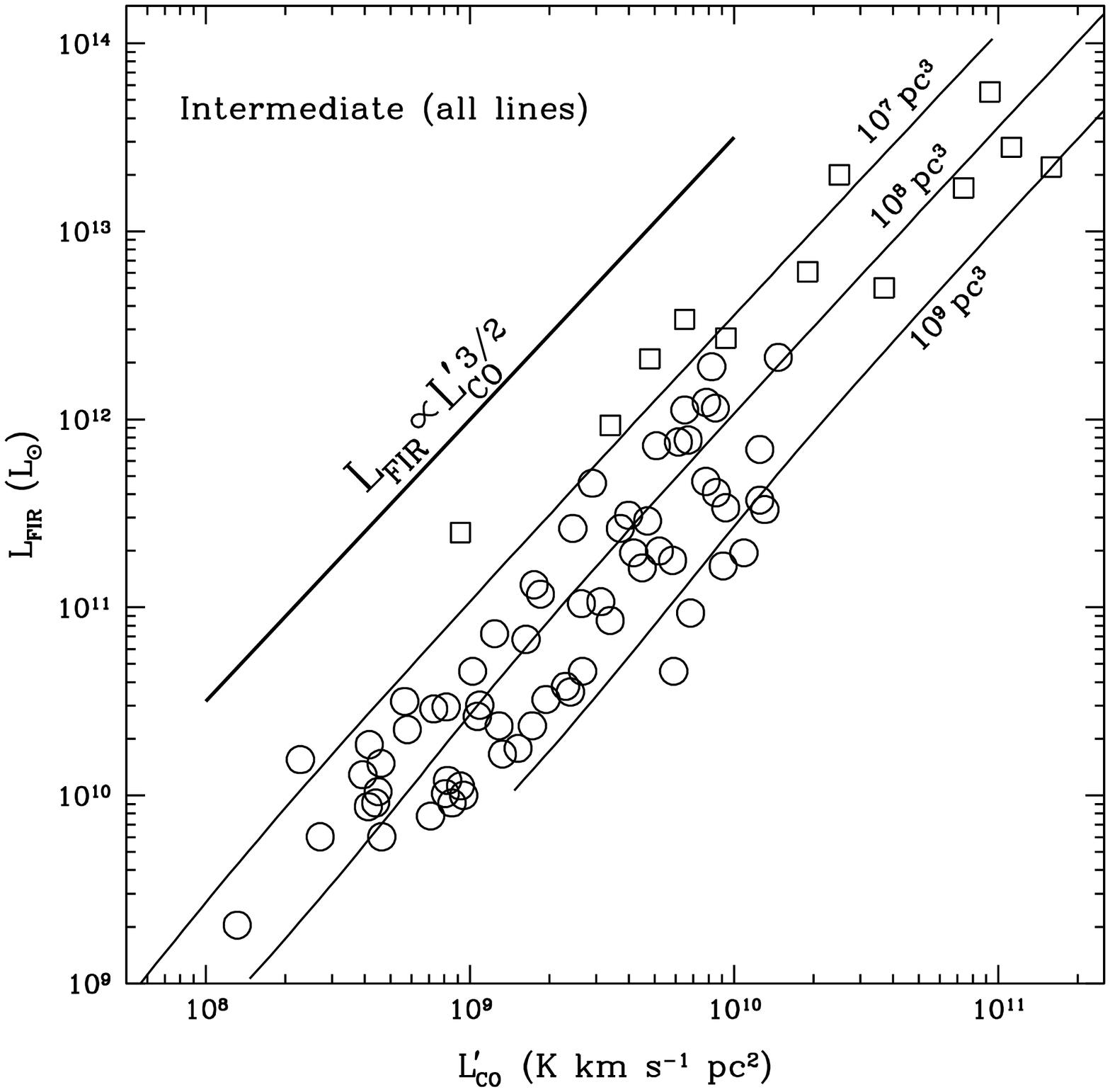}
\includegraphics[width=5.25cm]{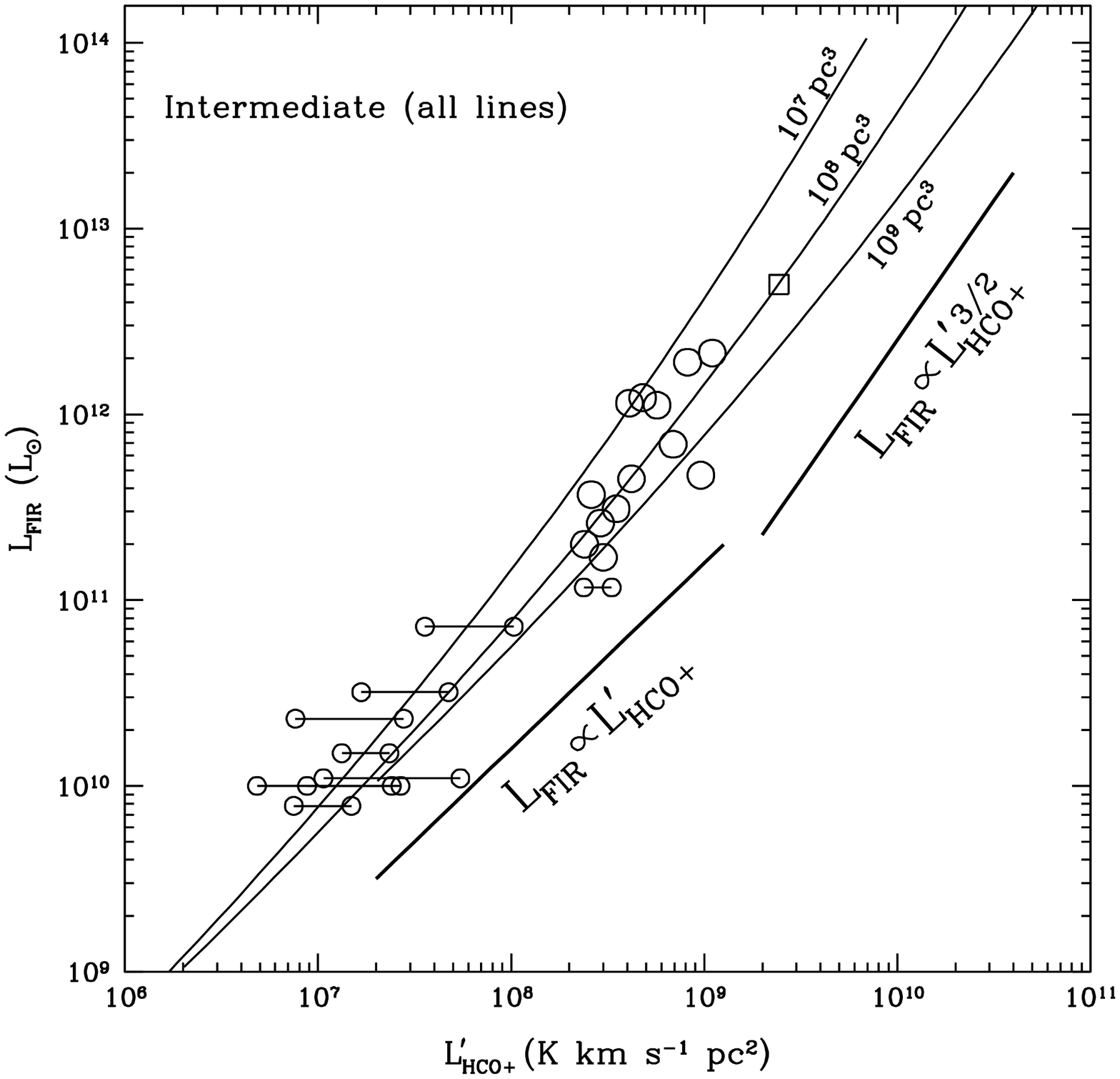}
\includegraphics[width=5.25cm]{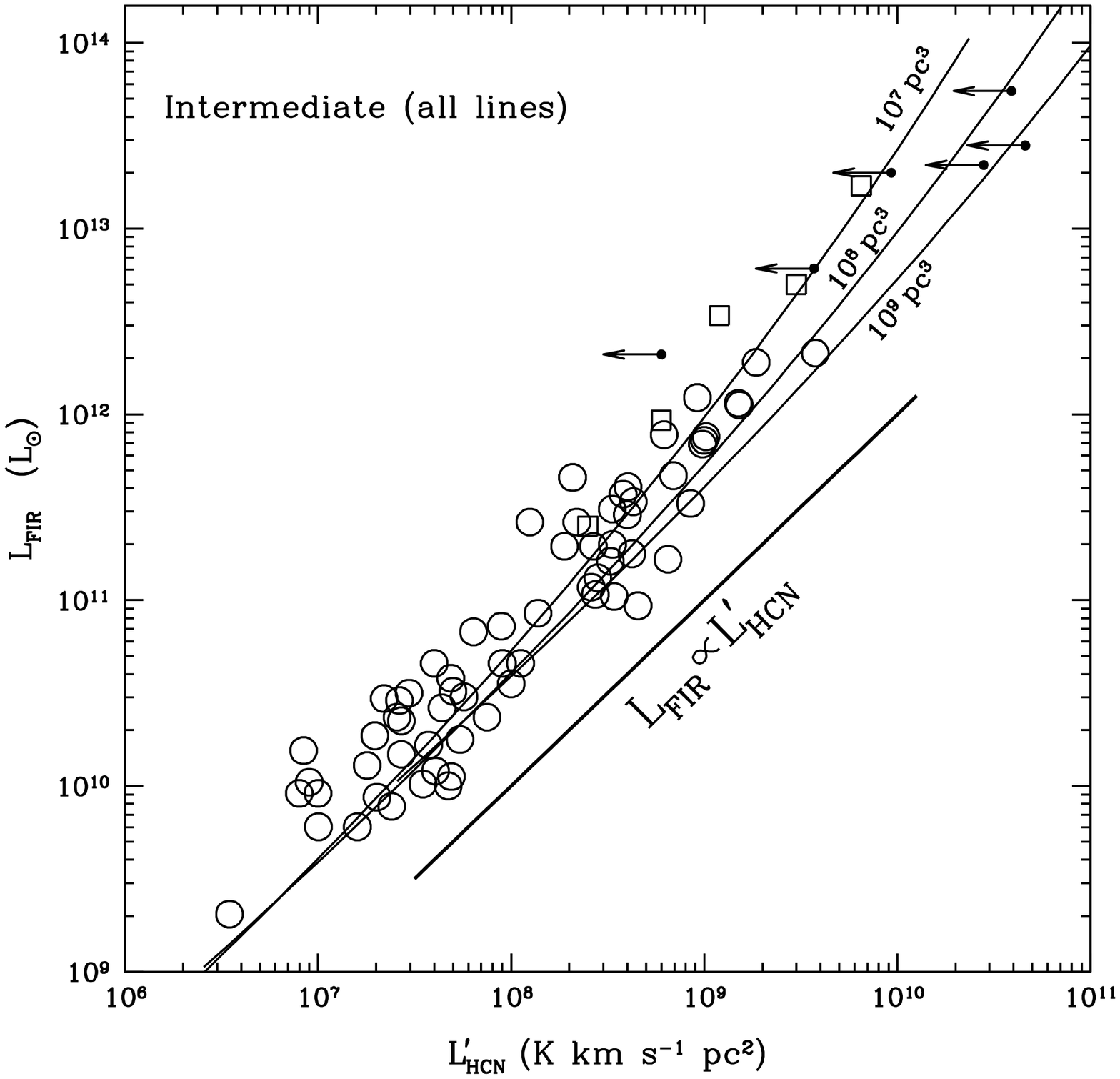} 
}
\figcaption[lums]{
$L_{\rm FIR}$ ($L_{\odot}$) versus $L_{\rm CO}^\prime$($\jline$)
({\it left panel}),  $L_{\rm HCO^+}^\prime$($\jline$) ({\it middle
panel}), and $L_{\rm HCN}^\prime$($\jline$) (K km s$^{-1}$ pc$^{2}$;
{\it right panel}). The lines in each panel derive from the
model presented in this paper with a constant total volume of
molecular material of $10^7$, $10^8$, and $10^9$ pc$^3$
(lowest to highest).  The thick solid
line segment shows power-law slopes to guide the eye. Data in the
left and right panels are from \citet{gao04b,gao04a} ({\it circles})
and \citet{gao07a} ({\it open squares} for detections,
{\it arrows} for upper limits).
The middle panel combines data from \citet{nguyen92}
({\it small circles with lines}), \citet{graciacarpio06} ({\it big
circles}), and \citet{riechers06b} ({\it open square}; using
the \citet{gao07a} FIR luminosity and magnification factor).
For all data, $L_{\rm FIR}$ is defined based on a weighted sum of the 
galaxy luminosity in the 60 and 100 $\mu$m IRAS bands, as described by 
\citet{sanders96}.
For the \citet{nguyen92} data, the uncertainties in $L_{\rm HCO^+}$
indicated by the lines arise because \citeauthor{nguyen92}\
provide both HCN($\jline$) and HCO$^+$($\jline$) intensities, but the
values for $L^\prime_{\rm HCN}$ derived from their work generally
fall a factor of $2-3$ below the $L^\prime_{\rm HCN}$ from
\citet{gao04b,gao04a} for the same systems. This is probably
because \citeauthor{nguyen92}\ use a single beam pointing rather
than integrating fully over extended sources, and therefore miss some
of the flux. We therefore show two
values of $L^\prime_{\rm HCO^+}$, connected by a line, for each
\citeauthor{nguyen92}\ data point: a smaller value calculated directly
from the data listed in their Table 2, and a larger value obtained by
multiplying the $L^\prime_{\rm HCN}$ value of \citeauthor{gao04b}
for that galaxy by the ratio $I_{\rm HCO+}/I_{\rm HCN}$ measured by
\citeauthor{nguyen92} If this ratio is constant over the source, this
estimate should correctly account for the flux outside the beam in the
\citeauthor{nguyen92}\ HCO$^+$ observation.\\
\label{fig:l}}
\end{figure*}

\begin{figure*}
\begin{center}
\includegraphics[width=5.25cm]{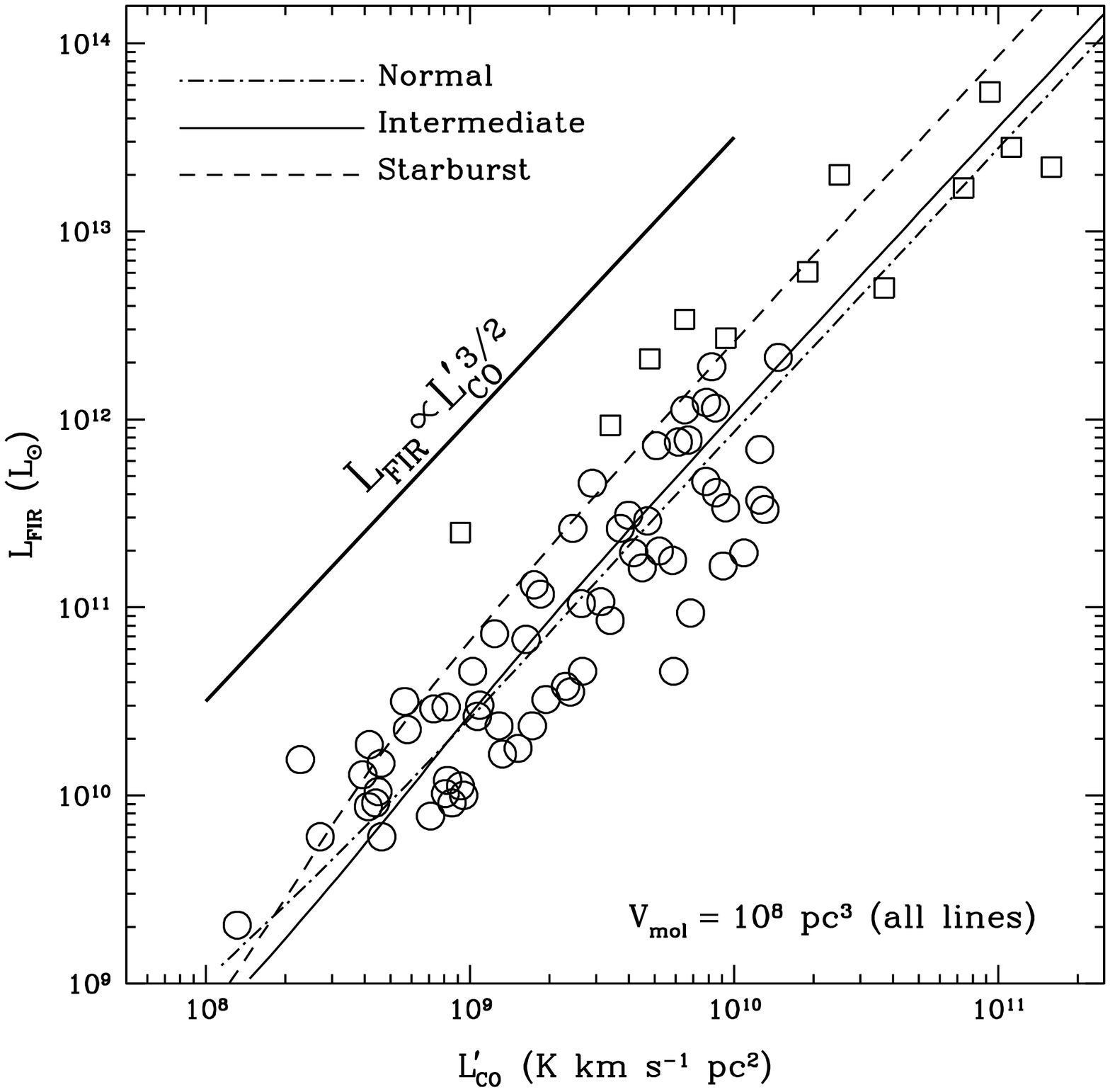}
\includegraphics[width=5.25cm]{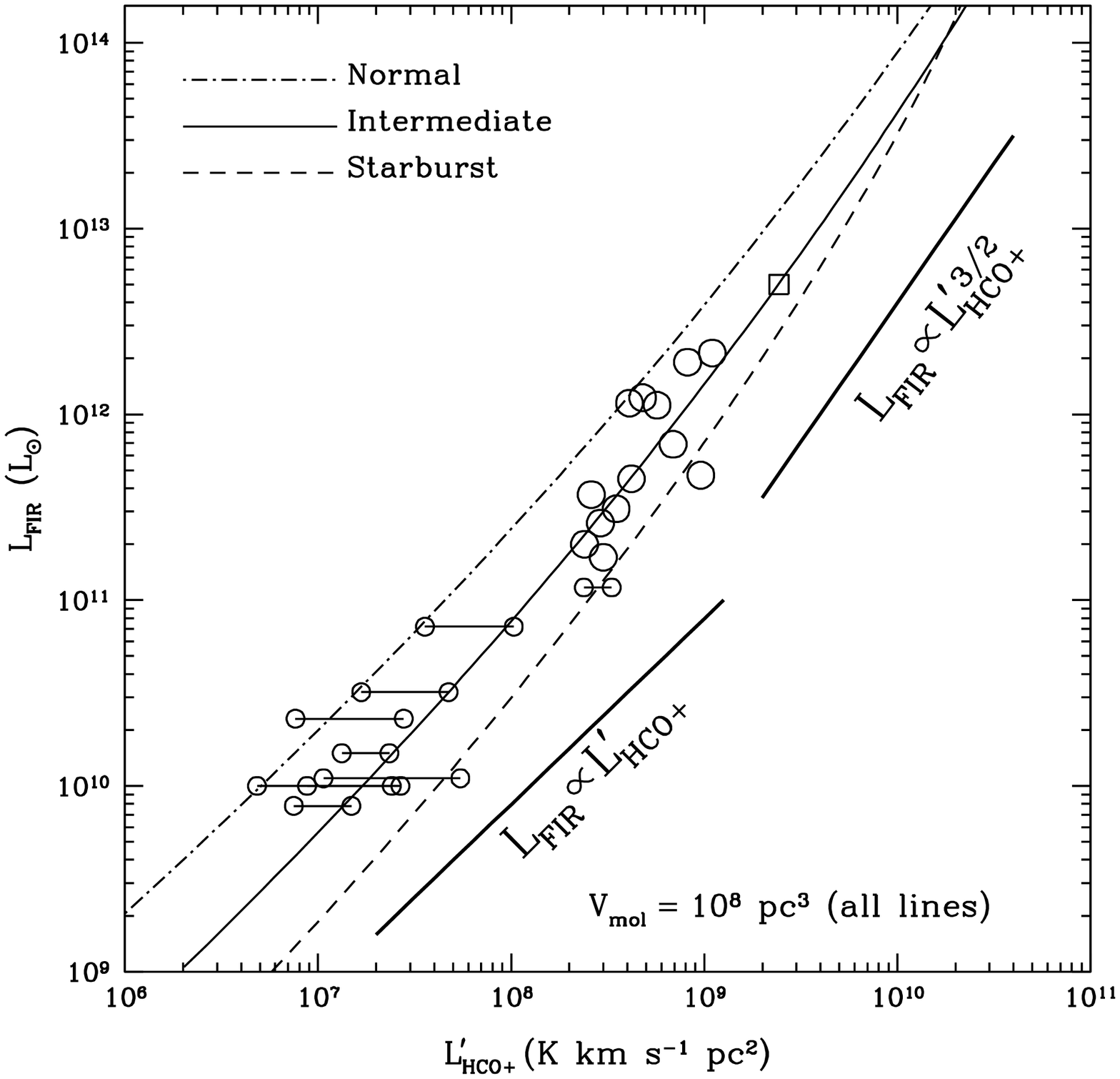}
\includegraphics[width=5.25cm]{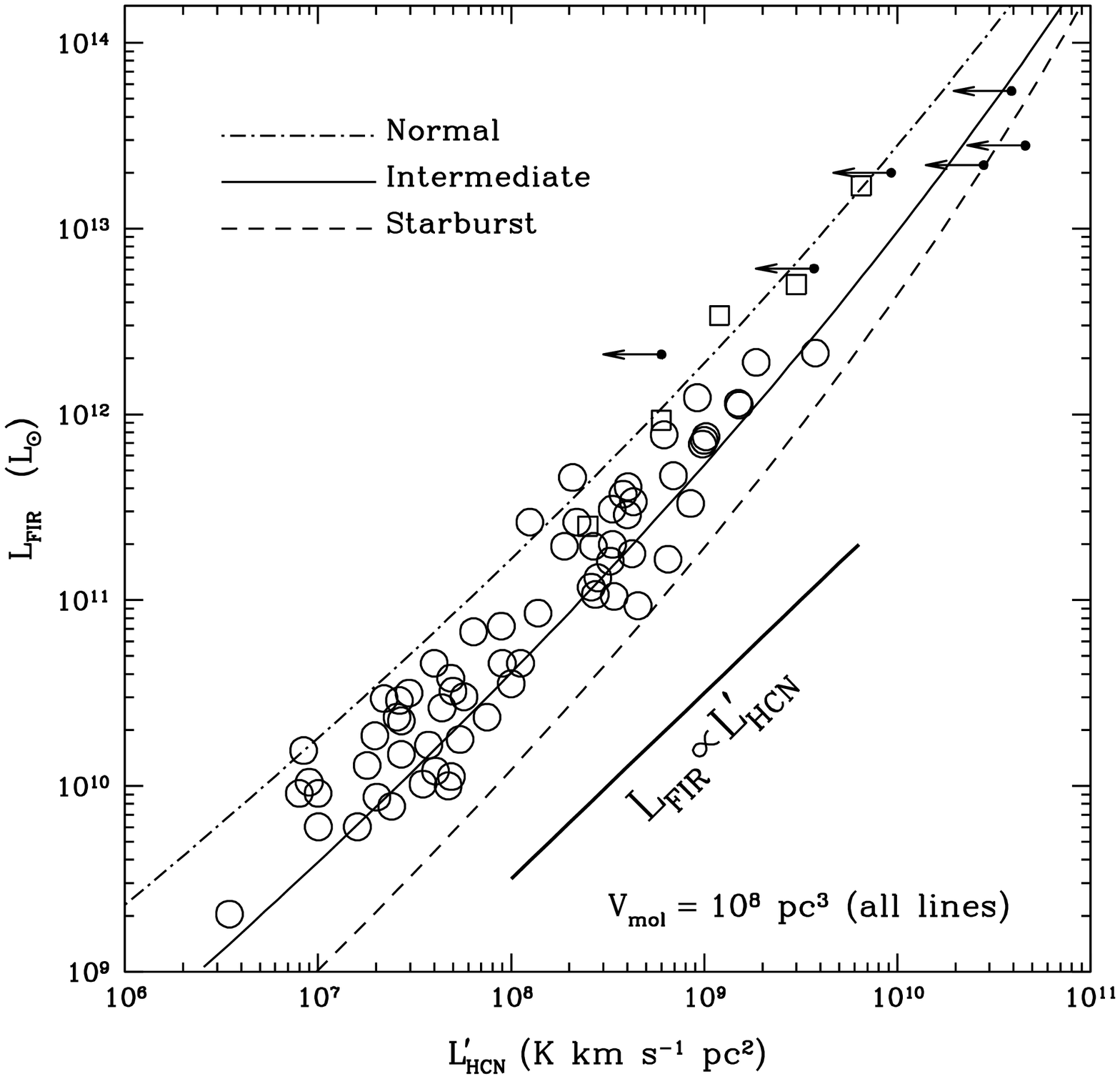} 
\end{center}
\figcaption[lumsm]{
The same as Figure \ref{fig:l}, but with constant
$V_{\rm mol}=10^8$ pc$^3$, and for the model parameters
corresponding to ``starburst'' ({\it dashed}), ``intermediate''
({\it solid}), and ``normal'' ({\it dot-dashed}) 
(see Table \ref{fiducial}).  Therefore, the middle
solid line in each panel of Figure \ref{fig:l} is the 
same as the solid line in each panel in this Figure.\\
\label{fig:lm}}
\end{figure*}

The calculations illustrated in Figure \ref{sfrl} demonstrate the
basic argument that one expects a roughly constant star formation rate
per unit line luminosity for high density tracers (e.g., HCN), and a star
formation rate per unit luminosity that rises like luminosity to the
$\sim 0.5$ for low density tracers (e.g., CO). However, in large surveys one
cannot always determine the mean density in a galaxy, which would be
required to construct an observational analog to Figure
\ref{sfrl}. Instead, we can use our calculated dependence of star
formation rate and line luminosity on density to compare to
observations as follows. Equation (\ref{linelum}) gives the total
molecular line luminosity per unit
volume and equation (\ref{SFR}) gives the star formation rate, which
we convert to an IR luminosity via equation (\ref{sfrtolir}).
For fixed assumed volume of molecular star-forming gas ($V_{\rm mol}$)
we can then  predict the expected correlations between $L^\prime$ in a
given molecular line  and $L_{\rm FIR}$.  The three panels of Figure
\ref{fig:l} show our results for $L_{\rm FIR}$ as a function of $L_{\rm
CO}^\prime$, $L_{\rm HCN}^\prime$, and $L_{\rm HCO^+}^\prime$ for 
the intermediate model (see Table \ref{fiducial}) and for several
values of $V_{\rm mol}$. 
Figure \ref{fig:lm} shows how are results vary as a function of the assumed
$T$ and $\calm$.  There, for fixed $V_{\rm mol}$, we compare our predictions
for the intermediate model with the normal and starburst models.
In both figures we compare our models to data culled from the 
literature.

From the work of \citet{gao04b,gao04a}, \citet[their Fig.~7]{greve05}, 
\citet[their Fig.~5]{riechers06b}, and \citet{gao07a},
as well as the theoretical
arguments in the preceding sections, we expect a strong, but
not linear, correlation between the CO luminosity and the star formation
rate --- as measured by $L_{\rm FIR}$ --- with the approximate form 
$L_{\rm FIR}\propto L'^{3/2}_{\rm CO}$. The left panel of 
Figure \ref{fig:l} shows the CO data, the approximate correlation 
expected ({\it solid line segment}; offset from the data for clarity) 
and the theoretical prediction ({\it solid lines})
for a total volume of molecular gas of $V_{\rm mol}=10^7$, $10^8$, and $10^9$ pc$^3$.
Because at fixed $L_{\rm FIR}$, galaxies exhibit a dispersion in
$V_{\rm mol}$ we expect there to be intrinsic scatter in this 
correlation, roughly bracketed by the range of $V_{\rm mol}$
plotted.

The middle and right panels of Figure \ref{fig:l} show the same prediction for $L_{\rm
HCO^+}^\prime$ and $L_{\rm HCN}^\prime$.  In these cases, because the 
molecular line luminosity is nearly linearly proportional to 
$L_{\rm FIR}$, the dependence on $V_{\rm mol}$ is much weaker
than for $L_{\rm CO}^\prime$.  However, systematic changes
or differences in the fiducial parameters for the calculation  
(see Table \ref{fiducial}) introduce uncertainty and scatter into 
the correlation.  Figure \ref{fig:lm} assesses this dependence.  It 
compares the predictions of our model for normal ({\it dot-dashed lines}), 
intermediate ({\it solid lines}), and starburst ({\it dashed lines})
galaxies, as defined in Table \ref{fiducial}, for fixed $V_{\rm mol}=10^8$ pc$^3$.
Our simple model reproduces the data rather well, and it predicts 
that generically there may be
more intrinsic scatter in the $L^\prime_{\rm CO}-L_{\rm FIR}$
correlation than in either $L^\prime_{\rm HCN}-L_{\rm FIR}$ or $L^\prime_{\rm
HCO^+}-L_{\rm FIR}$.

Note that in both the middle and right panels of Figures \ref{fig:l}
and \ref{fig:lm}, one expects a turn upward in the correlation at high 
$L_{\rm FIR}$, a deviation from linearity. This follows from the fact that
in our model, at fixed $V_{\rm mol}$, systems with higher $L_{\rm FIR}$
have higher average densities.  At sufficiently high $L_{\rm FIR}$ we
thus expect $L^\prime_{\rm HCN}-L_{\rm FIR}$ and $L^\prime_{\rm HCO^+}-L_{\rm FIR}$
to steepen, in analogy with the $L^\prime_{\rm CO}-L_{\rm FIR}$ correlation.  
The data points with very high $L_{\rm FIR}$ in Figures \ref{fig:l}
and \ref{fig:lm}, which might be used to test this prediction of our model,
are gravitationally lensed, at high redshift, and contaminated by bright AGN.  
It is therefore unclear if the deviation from 
linearity implied particularly by the upper limits in $L^\prime_{\rm HCN}$
in the right panels of Figures \ref{fig:l} and \ref{fig:lm} is a result
of enhanced $L_{\rm FIR}$, caused by the AGN emission \citep{carilli05},
or is instead a result of less molecular line emission per unit star formation,
as our model implies (Fig.~\ref{sfrl}).  \citet{gao07a} note, however, 
that in the three systems for which the contribution
from the AGN has been estimated (F10214+4724, D.~Downes \& P.~Solomon 2007,
in preparation; Cloverleaf, \citealt{weiss03}; APM 08279+5255, \citealt{weiss05,weiss07}) 
the corrections are only significant for APM 08279+5255.  This 
suggests that the data are so far consistent with our interpretation,
but clearly much more data at high $L_{\rm FIR}$  --- or, more 
precisely for our purposes, at high density --- is required to 
test our predictions. We discuss the issue of AGN contamination 
further in \S~\ref{limitations}.


As a final note, the data so far {\it do} support the utility of
HCO$^+$ as a useful tracer of dense gas. \citet{papadopoulos07} has
argued against the utility of HCO$^+$ as a faithful tracer of mass in
starbursts on the basis that, since it is an ion, its abundance is
strongly dependent on the free-electron abundance and might therefore
vary strongly between galaxies with different ionizing radiation
backgrounds. We cannot rule out this possibility given the limited
data set available, but we see no strong evidence in favor of it from
the data shown in Figures \ref{fig:l} and \ref{fig:lm}.  As we have
argued, HCO$^+$($\jline$) is particularly 
useful because its critical density is between that of CO($\jline$) and
HCN($\jline$) and, thus, as Figure \ref{fig:l} and \ref{fig:lm} show,
the correlation between $L_{\rm FIR}$ and $L^\prime_{\rm HCO+}$ should
steepen from linear to super-linear over the range of galaxies presented
in the CO panels.  A careful, large-scale HCO+($\jline$) survey 
similar to the work of \citet{gao04b} on HCN($\jline$) should reveal
these trends.  Lines with similarly low excitation temperatures and
intermediate critical densities like CS($\jline$) should behave
analogously.

\section{Discussion}
\label{discussion}

\subsection{Implications for Kennicutt-Schmidt Laws and Star Formation
Efficiencies}

Our results suggest that KS laws in different tracers
naturally fall into two regimes, although there is a broad range of
molecular tracers that are intermediate between the two
extremes. Tracers for which the critical density is small compared to
the median density in a galaxy represent one limit. In these tracers,
the light faithfully follows the mass, so the KS law
measures a relationship between total mass and star formation. In any
model in which star formation occurs at a roughly constant rate per
dynamical time, this must produce a KS law in which the
star formation rate rises with density to a power of near $1.5$, and the
ratio of star formation to luminosity rises as density to the $0.5$
power. In terms of surface rather than volume densities, this implies
$\dot{\Sigma}_* \propto \Sigma_g^{3/2} h^{-1/2}$. If we further add
the observation that the scale heights $h$ of the star-forming
molecular layers of galaxies are roughly constant across galaxy types,
one form of the observed \citet{kennicutt98b,kennicutt98a} star
formation law follows immediately \citep{elmegreen02}. Moreover, in a
galactic disk, $h\propto \Sigma_g/\nbar$ and $\nbar\propto \Omega^2/Q$
\citep[e.g.][]{thompson05}; since in star-forming disks the Toomre-$Q$
is about unity \citep{martin01}, substituting for $h$ immediately
gives $\dot{\Sigma}_*\propto \Sigma\Omega$, the alternate form of
the \citet{kennicutt98b,kennicutt98a} law. 

The other limit is tracers for which the
critical density is large compared to the median galactic
density. These tracers pick out a particular density independent of
the mean or median density in the galaxy, and thus all the regions
they identify have the same dynamical time regardless of galactic
environment. In this case the star formation rate will simply be
proportional to the total mass of the observed regions, yielding a
constant ratio of star formation rate to mass, as is observed for HCN
in the local universe \citep[Fig.~\ref{fig:l}, right
panel;][]{gao04b,gao04a,wu05}.

We predict that there should be a transition between linear and
super-linear scaling of $L_{\rm FIR}$ with line luminosity at the point
where galaxies transition from median densities that are
smaller than the line critical density to median densities larger than
the critical density. The HCO$^+$($\jline$) line, and other lines with 
similar critical densities, e.g. CS($\jline$) and SO($\jline$), should 
show this behavior for galaxies in the local universe. The observed correlation 
between $L_{\rm HCO^+}$ and $L_{\rm FIR}$ appears to be consistent with our
prediction, although at present the data are not of sufficient quality to
distinguish between a break and a single powerlaw relation. There are
hints that the very highest luminosity star-forming galaxies, which
all reside at high redshift and may well reach ISM densities not found
in any local systems, show such a break in the IR-HCN correlation. 

One important point to emphasize in this analysis is that we have been
able to explain the observed correlations between line and infrared
luminosities, and hence between gas masses at various densities and
star formation rates, without resorting to the hypothesis that the
star formation process is fundamentally different in galaxies of
different properties. 
Although uncertainties in both our model and 
the observations do not preclude an order-unity change in the
star formation efficiency or $\sfrff$ as a function of $L_{\rm FIR}$,
there is currently no evidence for such a change in the data, contrary
to claims made by, e.g.\ \citet{graciacarpio06}.
In fact, all of the observational trends are predicted by our simple model with 
{\it constant} star formation efficiency. This is consistent with other lines of 
evidence that the fraction of mass at a given density that turns into stars 
is roughly 1\% per free-fall time independent of density \citep{krumholz07e}.

\subsection{Does Star Formation Have a Fundamental Size or Density Scale?}

Based on the linear correlation between HCN($\jline$) luminosity and
star formation rate, seen both in external galaxies and in individual
molecular clumps in the Milky Way, \citet{gao04b,gao04a} and
\citet{wu05} propose that HCN($\jline$) emission traces a fundamental
unit of star formation. They explain the linear IR-HCN correlation as
a product of this; in their model, HCN luminosity correlates linearly
with star formation rate because HCN luminosity simply counts the
number of such units.

Based on our analysis, we argue that this hypothesis is only partially
correct. We concur with Gao \& Solomon and Wu et al.\ that the
HCN($\jline$) luminosity of a galaxy does simply reflect the mass of
gas that is dense enough to excite the HCN($\jline$) line. However,
our analysis shows that this does not necessarily imply that this
density represents a special density in the star formation process, or
that objects traced by HCN($\jline$) represent a physically distinct
class. We show that a linear correlation between star formation rate
and line luminosity is
expected for any line with a critical density comparable to or larger
than the median molecular cloud density in the galaxies used to define
the correlation.
It is possible that HCN($\jline$)-emitting
regions represent a physically distinct scale of star formation as
\citeauthor{wu05} propose, but one can explain the linear IR-HCN
correlation equally well if they are just part of the same
continuous medium as the regions traced by CO($\jline$) and by other
transitions. Even the star-forming clouds themselves may simply be parts
of a continuous distribution of ISM structures occupying the entire
galaxy, as argued by \citet{wada07}. In this case there need be no
special density scales other than the mean and median densities for
the star-forming clouds on their largest scales, and the density at
which star formation becomes rapid, converting the mass into stars in
of order a free-fall time. This transition
scale is unknown, but must be considerably larger than the density
traced by HCN \citep{krumholz07e}.

\subsection{Limitations and Cautions}
\label{limitations}

\subsubsection{Self-Consistency}

As mentioned in \S~\ref{schmidtlaws}, our approach of 
leaving the gas temperature $T$ and Mach number $\calm$ as free parameters 
is not entirely consistent with our calculation of the IR luminosity, 
since the IR luminosity and temperature are of course related. In 
principle, with a model of how the energy output from stars heats 
the dust and gas, together with a structural model connecting the
energy and momentum output from stars to the generation of turbulence,
it should be possible to self-consistently compute both the gas 
temperature and the Mach number from the volumetric star formation 
rate \citep[see, e.g.,][]{thompson05}.  
Such a model would return $T$ and $\calm$ as a function of 
$\nbar$ and possibly other galaxy properties, while simultaneously
predicting a set of Kennicutt-Schmidt laws.  

If the line luminosity depended strongly on $T$ or $\calm$, or if one 
required knowledge of the temperature to compute the infrared luminosity 
of a galaxy, we would would have no alternative to constructing such a 
model if we wished to explain the observed IR-line luminosity correlation. 
However, we can avoid this by relying on the observationally-calibrated 
star formation-IR correlation, and because, as we show in Figures 
\ref{sfrl}, \ref{fig:l}, and \ref{fig:lm}, the line luminosity varies quite 
weakly over a reasonable range of $T$ and $\calm$ for our chosen lines. 
For this reason, any model for computing $T$ and $\calm$ as a function of 
galaxy properties, if it were consistent with observations, would not 
significantly alter the IR-line luminosity correlation we derive. This is 
true, however, only for lines that require low temperatures to excite. 
As we discuss in \S~\ref{isothermality}, lines that require higher
temperatures to excite do depend sensitively on the temparature in the
galaxy, and a model capable of predicting the IR-line luminosity
correlation for these lines must also include a calculation of the
temperature structure of the galaxy.

\subsubsection{Isothermality}
\label{isothermality}

Our assumption of isothermality means that our
analysis will only apply to molecular lines for which the temperature 
$T_{\rm up}$ corresponding to the upper state energy is $<10$ K, low 
enough to be excited even in the coolest
molecular clouds in normal spiral galaxies. The reason for this is
that at temperatures larger than $T_{\rm up}$, the
luminosity in a line generally varies at most linearly with the
temperature. As the similarity between the results with our different 
galaxy models illustrates, changing the temperature within the range 
of $\sim 10-50$ K produces only a factor of a few change in the luminosity 
of the lines we have studied. In contrast, line luminosity responds
exponentially to temperature changes when the temperature is below the 
value corresponding to the upper state's energy. This means that lines 
sensitive to high temperatures pick out primarily the regions that are warm 
enough for the line to be excited. Density has only a secondary effect. The 
emission will therefore reflect the temperature distribution in star-forming clouds
more than the density distribution, an effect that our isothermal assumption
precludes us from treating. KS laws in high temperature tracers
are likely to find linear relationships between star formation rate
and mass regardless of the critical density of the molecule in
question because they will simply be correlating the mass of dust warmed 
to $\gtsim 100$ Kelvin, which is essentially what is measured by $L_{\rm FIR}$, 
with the mass of gas warmed to temperatures above $T_{\rm up}$. However, our 
model will not apply to these lines, and for this reason we do not attempt to 
compare to observations using higher transitions of CO ($3\rightarrow 2$, 
$4\rightarrow 3$, $5\rightarrow 4$, $6\rightarrow 5$, and $7\rightarrow 6$, 
which have $T_{\rm up} = 33, 55, 83, 116$, and $154$ K, respectively; 
\citealt{greve05}, \citealt{solomon05}), CS($5\rightarrow 4$) 
($T_{\rm up}=35$ K; \citealt{plume97}), or other high temperature tracers.

\subsubsection{Molecular Abundances}

We have not considered
density-dependent variations in molecule abundances. One potential
source of variation in molecular abundance is freeze-out onto grain
surfaces at high densities and low temperatures
\citep[e.g.][]{tafalla04a, tafalla04b}. Chemodynamical models suggest
that freeze-out is not likely to become significant for either
carbonaceous or nitrogenous species until densities $n\gtsim 10^6$
cm$^{-3}$ \citep{flower06}, but may become severe at higher densities,
so whether depletion is significant depends on what fraction of the
total luminosity would be contributed by gas of this density or higher
were there no freeze-out. Figure \ref{dldlogn} suggests that
freeze-out is likely to modify the total galactic luminosity of CO,
HCO$^+$, and HCN fairly little even at a mean ISM density of $\nbar=10^4$
cm$^{-3}$, but may have significant effects for galaxies of
larger mean densities or for lines for which the critical densities is
comparable to the freeze-out density.
If freeze-out is significant, our conclusions will be modified.

\subsubsection{Atomic Gas}

In the simple model developed here, we have neglected the role of atomic gas 
entirely. Whether the density or surface density of atomic gas plays 
a role in controlling the star formation rate is subject to debate on 
both observational and theoretical grounds 
\citep{kennicutt98b,kennicutt98a,wong02,heyer04b,komugi05,krumholz05c, kennicutt07a}, so it 
is unclear how much a limitation this omission really is. We can say with 
confidence that in molecule-rich galaxies, which provide almost all the 
dynamic range of both the \citet{kennicutt98b,kennicutt98a} correlation 
and the correlations illustrated in Figures \ref{fig:l} and
\ref{fig:lm}, the atomic gas  plays almost no role simply because
there is so little of it. Thus, our  predictions should be quite
robust, except perhaps at the very low luminosity ends of Figures
\ref{fig:l} and \ref{fig:lm}.

\subsubsection{AGN Contributions}
\label{agn}

A final point is not so much a limitation of our work as a cautionary 
note about comparing our model with observations. We have included in our 
model IR luminosity only from star formation, and molecular line luminosity 
only from molecules in cold star-forming clouds. However, an AGN may make a 
significant contribution to a galaxy's luminosity in the far infrared by 
direct heating of dust grains, and in molecular lines via an X-ray dissociation 
region.  Indeed, several of the systems with the highest IR luminosities in
Figures \ref{fig:l} and \ref{fig:lm} are contaminated by AGN.  As noted in 
\S\ref{obscomparison}, this complicates an assessment of our prediction of 
an up-turn in the $L^\prime_{\rm HCN}-L_{\rm FIR}$ and $L^\prime_{\rm HCO+}-L_{\rm FIR}$
correlations at high luminosity.  This deviation from linearity at high gas density
(at fixed $V_{\rm mol}$, high $L_{\rm FIR}$) is an essential prediction of 
our model, but testing it relies on a careful separation of the 
contribution of the AGN to both the IR and line luminosities \citep[e.g.][]{maloney96}.
In fact, \citet{carilli05} discuss the possibility that the AGN's contribution
to the IR luminosity in these systems causes them to be
above the local linear $L^\prime_{\rm HCN}-L_{\rm FIR}$ correlation.  
Such a contamination would mimic the prediction of our model.
However, \citet{gao07a} argue that the sub-millimeter galaxies in
their sample are not AGN dominated and that just one of three 
quasars in their sample (APM 08279+5255) has a large AGN IR
component.   See \citet{gao07a} for more discussion.  
For these reasons we contend that although our model is 
consistent with the existing data, the current evidence for 
a break in the $L^\prime_{\rm HCN}-L_{\rm FIR}$ correlation 
should be viewed with caution and more data in high density/luminosity
systems is clearly required to understand the role of AGN
contamination in shaping the correlation.

\section{Conclusions}
\label{conclusions}

We provide a simple model for understanding how Kennicutt-Schmidt
laws, which relate the star formation rate to the mass or surface
density of gas as inferred from some particular line, depend on the
line chosen to define the correlation. We show that for a turbulent
medium the luminosity per unit volume in a given line, provided that 
line can be excited at temperatures lower than the mean temperature 
in a galaxy's molecular clouds, increases
faster than linearly with the density for molecules with critical
densities larger than the median gas density. The star formation rate
also rises super-linearly with the gas density, and the combination of
these two effects produces a close to linear correlation between star
formation rate and line luminosity. In contrast, the line luminosity
rises only linearly with density for lines with low critical
densities, producing a correlation between star formation rate and
line luminosity that is super-linear.

Based on this analysis, we construct a model for the correlation between a galaxy's infrared luminosity and its luminosity in a particular molecular line. Our model is extremely simple, in that it relies on an observationally-calibrated IR-star formation rate correlation, it treats molecular clouds as having homogenous density and velocity distributions, temperatures, and chemical compositions, and it only very crudely accounts for variations in molecular cloud properties across galaxies. Despite these approximations, the model naturally explains why some observed correlations
between infrared luminosity and line luminosity in galaxies are
linear, and some are super-linear. Using it, we are
able to compute quantitatively the correlation between infrared and
HCN($\jline$) line luminosity, and between IR and CO($\jline$) line
luminosity. We show that our model provides a very good fit to
observations in these lines, and we are able to make similar
predictions for any molecular line that can be excited at low temperatures,
as we demonstrate for the example of HCO$^+$($\jline$). Moreover, we
are able to explain the observed data without recourse to the
hypothesis that the star formation process is somehow different,
either more or less efficient, in different types of
galaxies or for media of different densities. Instead, our model is
able to explain the observed correlations using a simple, universal
star formation law.

One strong prediction of our model is that there should be a break
from linear to non-linear scaling in the HCN-IR correlation at very
high IR luminosity, and a similar break in the HCO$^+$-IR correlation
at somewhat lower luminosity. The data for HCO$^+$ are consistent with
this prediction but do not yet strongly favor a break over pure
powerlaw behavior. However, there is some preliminary evidence for a
break in the IR-HCN correlation in high redshift galaxies more
luminous than any found in the local universe, although with these high 
redshift observations it is difficult to rule out the alternative explanation 
of the break as arising due to a progressively rising AGN contribution to the 
IR luminosity (see \S\ref{obscomparison} and \S\ref{agn}). Future galaxy surveys
both in the local universe and at high redshift may be used to test
our predictions for HCO$^+$($\jline$), HCN($\jline$), and other
molecular lines.

\acknowledgements 
We thank L.\ Blitz, B.\ Draine, A.\ Leroy,
E.\ Rosolowsky, and A.\ Socrates for helpful discussions, N.\ Evans 
and the anonymous referee for useful comments on the manuscript, and 
R.\ Kennicutt for kindly providing a preprint of his submitted paper. 
We thank Y.~Gao for providing $L_{\rm FIR}$ for the systems used in 
Figures \ref{fig:l} and \ref{fig:lm}.  
MRK acknowledges support from NASA through Hubble
Fellowship grant \#HSF-HF-01186 awarded by the Space Telescope Science
Institute, which is operated by the Association of Universities for
Research in Astronomy, Inc., for NASA, under contract NAS 5-26555. TAT
acknowledges support from a Lyman Spitzer, Jr.\ Fellowship.


\end{document}